\documentclass[runningheads]{llncs}
\usepackage[T1]{fontenc}
\usepackage{graphicx}
\usepackage{amsmath}

\begin{document}
\title{Comparing User Behavior in Real vs. Virtual Supermarket Shelves: An Eye-Tracking Study Using Tobii 3 Pro and Meta Quest Pro}
\titlerunning{Comparing User Behavior in Real vs. Virtual Supermarket Shelves}
%
\author{Francesco Vona\inst{1}\orcidID{0000-0003-4558-4989} \and
Julia Schorlemmer\inst{1}\orcidID{0009-0004-7388-9389} \and
Paulina Kaulard \inst{1}\orcidID{0009-0004-8686-8333} \and
Sebastian Fischer\inst{1}\orcidID{0000-0002-9711-0382} \and 
Jessica Stemann\inst{1}\orcidID{0000-0002-0361-6754} \and \\
Jan-Niklas Voigt-Antons\inst{1}\orcidID{0000-0002-2786-9262}}
\authorrunning{F. Vona et al.}
%
\institute{University of Applied Sciences Hamm-Lippstadt, Germany\\
\email{name.lastname@hshl.de}}
\maketitle              
\begin{abstract}
This study compares user behaviors between real and virtual supermarket shelves, using eye-tracking technology to assess behavior in both environments. A sample of 29 participants was randomly assigned to undergo two conditions: a real-world supermarket shelf with Tobii eye-tracking and a virtual shelf using Meta Quest Pro’s eye-tracker. In both scenarios, participants were asked to select three packs of cereals belonging to specific categories (healthy/tasty). The aim was to explore whether virtual environments could realistically replicate real-world experiences, particularly regarding consumer behavior. By analyzing eye-tracking data, researchers examined how attention and product selection strategies varied between real and virtual conditions. 
Participants' attention differed across product types and shopping environments. Consumers focus on lower shelves in real settings, especially when looking for healthy products. In VR, attention shifted to eye-level shelves, particularly for tasty items, aligning with optimal product placement strategies in supermarkets. Generally, sweet products received less visual attention overall.
\end{abstract}
\section{Introduction}
With the advancement of Virtual Reality (VR), consumer behavior research has gained a powerful tool for simulating shopping experiences \cite{Branca2023-yf}. VR enables researchers to control variables in lifelike yet controlled environments, allowing for precise studying of behavioral responses. Through VR, researchers can evaluate how consumers react to store layouts, product placements, and packaging designs without requiring a physical test location. Moreover, VR captures detailed eye-tracking and spatial data, providing real-time insights into attention, engagement, and decision-making processes \cite{Branca2024-ef}.
Eye-tracking technology in VR offers a valuable perspective on consumer attention and interaction with products. By analyzing where and how long consumers focus on certain items, researchers can identify which products draw initial attention, sustain engagement, or influence purchase intentions. This data is particularly relevant in retail settings, where product placement—such as positioning at eye level versus lower shelves—can significantly impact purchasing decisions. Eye-tracking allows businesses to optimize shelf arrangements and promotional displays based on objective, behavior-driven insights \cite{bettiga2020consumer,huddleston2015consumer,MEINER2019445}.

\noindent
In this study, we examine the effect of virtual environments on consumer behavior by comparing interactions in real and virtual supermarket settings. To achieve this, we recreated a supermarket shelf in both a physical location and a VR environment, allowing participants to explore different cereal categories. Using eye-tracking technology, we analyzed consumer attention patterns and preferences across different product types.
By collecting and comparing eye-tracking data, we investigate whether consumer attention varies across product categories, providing insights into decision-making tendencies and product preferences. For instance, consumers may spend more time examining health-focused cereals than sweeter or more indulgent options. Our study explores how these visual attention patterns differ between real and virtual settings, highlighting the potential of VR as a research tool for consumer behavior and retail optimization.

\section{Related Work}
The use of Virtual Reality (VR) in consumer behavior research has demonstrated the potential of virtual environments to closely replicate real-world shopping experiences. Studies have validated VR as a realistic simulation tool, showing that consumer behavior in virtual shopping settings largely mirrors that in physical stores. For example, research comparing consumer interactions in VR supermarket shelves to real stationary shelves of similar dimensions found that behavior in VR does not significantly differ from that in physical retail environments, reinforcing VR’s reliability in replicating real-world shopping experiences \cite{Siegrist2019-yo}. Similarly, studies exploring product packaging perception found minimal differences between consumer evaluations of virtual and real product presentations \cite{Branca2023-yf}. These results highlight VR’s ability to simulate shopping environments effectively, though factors such as the range of available products can influence the overall consumer experience. Additional research suggests that a limited product selection in VR may reduce user satisfaction, emphasizing that product variety is crucial to enhancing the VR shopping experience \cite{Pizzi2019-ei}.
Beyond assessing VR’s realism, other studies have utilized virtual environments to examine factors influencing consumer attention and decision-making \cite{Schnack_2020}. Research conducted in controlled virtual supermarkets confirms the importance of shelf placement in guiding consumer attention, with products positioned at eye level receiving greater focus and generating higher purchase intent \cite{Branca2023-yf,Siegrist2019-yo}. These findings reinforce the long-established retail concept of strategic product placement, now validated in VR, where variables can be meticulously controlled \cite{vanHerpen2016}.
Expanding on the psychological aspects of consumer interactions in digital spaces, researchers have applied assemblage theory to study how immersive VR and AR settings affect perceptions of product ownership and permanence \cite{hoffman2018iot}. Findings suggest that the unique characteristics of VR environments can shape how consumers evaluate and relate to products, reinforcing VR’s potential to create emotionally engaging shopping experiences.
Personalized advertising in VR environments has also gained attention. Research suggests that contextually relevant, tailored advertising within VR enhances brand attitudes by making brand interactions feel more authentic and engaging \cite{verhulst2017adcontext}. These results underscore the potential of VR and AR not only as experimental research tools but also as effective platforms for delivering customized marketing experiences.
Recent studies have further explored sensory and experiential dimensions of VR. Research examining the role of ambient scent in VR environments demonstrates how sensory cues can enhance emotional engagement and user satisfaction \cite{Helmefalk2019}. Additional research into digital sensory marketing emphasizes VR’s effectiveness in creating sensory-rich experiences that mimic real-world shopping \cite{Petit2018}. Other studies exploring technological embodiment in VR show that immersive devices increase user presence and strengthen behavioral intentions, suggesting that VR provides a more authentic and engaging consumer experience \cite{Flavian2019}.
Studies on digital elements in retail further highlight VR’s role in shaping consumer attention and product perception. Research into digital signage in VR retail settings shows that interactive displays and digital interfaces effectively guide consumer focus and replicate in-store cues \cite{Dennis2014}. Additional studies discuss how VR and AR technologies respond to growing consumer expectations for immersive, personalized shopping experiences, highlighting VR’s ability to balance entertainment and practicality in retail environments \cite{Grewal2017}.
Building on these findings, the present study seeks to compare consumer behavior in a real supermarket setting versus a virtual replica in VR, integrating an immersive supermarket environment.
\section{Methods}
\subsection{Experiment Design}
\begin{figure}[ht]
    \centering
    \includegraphics[width=\linewidth]{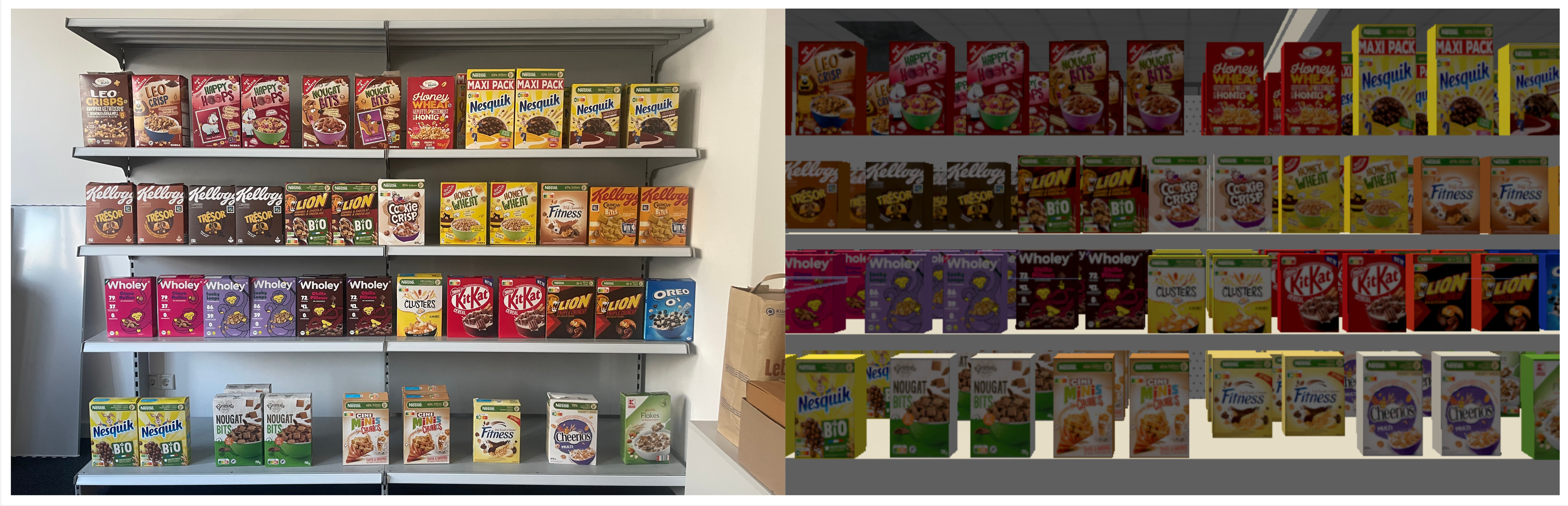}
    \caption{The figure shows a side-by-side comparison between a real supermarket shelf display (left) and a virtual representation in a VR environment (right).}
    \label{fig:realvsvirtulashelf}
\end{figure}

\noindent This study employed a within-subjects design to compare consumer behavior in two distinct shopping environments: a real supermarket shelf (real environment condition) and a virtual supermarket shelf (virtual environment condition) (Figure \ref{fig:realvsvirtulashelf}). Both environments featured an identical selection of 26 cereal products (SKUs), with each SKU appearing twice on the shelf (two facings). Participants completed tasks in which they selected three cereal packages from one of two predefined categories: “healthy” or “tasty.”
The real environment condition was conducted using a physical supermarket shelf with actual cereal packages, set up in a controlled laboratory setting at the university. The virtual environment condition was presented in a separate laboratory space, simulating a comparable virtual supermarket shelf displaying the same cereal products.
To control for order effects, participants experienced both environments in randomized order. In the real environment, participants wore Tobii Pro Glasses 3 to track their eye movements, while in the virtual environment, eye-tracking was recorded using the Meta Quest Pro’s integrated eye-tracker. Unique participant IDs were assigned, and task sequences within each condition were randomized to minimize potential biases.
\subsection{Participants}
A total of 39 participants were recruited for the study. However, 10 participants were excluded due to technological malfunctions resulting in missing data for at least one of the experimental tasks. The final sample consisted of 29 participants (74\% female). Participants were between 18 and 30 years old, with a mean age of 20.44 years (SD = 2.27). Before participation, all individuals were informed about the study’s purpose and provided written consent in accordance with ethical guidelines.
\subsection{Procedure}
The experiment took place in a controlled laboratory setting on the university campus following ethical approval from the university’s ethics committee. Each session was conducted individually and lasted approximately 20 minutes. Two adjacent rooms in the psychology laboratory were designated for the two experimental conditions. One room contained the real supermarket shelf, while the other room was used to present the virtual environment. Upon arrival, participants were greeted by two researchers, who provided an overview of the study and obtained informed consent via a signed form. Before beginning the tasks, participants completed a preliminary questionnaire collecting demographic information.
Participants were randomly assigned to start in either the real or virtual environment.
In each environment, participants were fitted with either the Tobii eye-tracking glasses (real environment) or the Meta Quest Pro HMD (virtual environment). Once the devices were properly fitted, the researchers provided task instructions. Participants were asked to select three cereal packages from their assigned category (“healthy” or “tasty”) and place them in a designated shopping container. In the real environment, selections were placed in a physical basket, while in the virtual environment, selections were placed in a virtual shopping cart.
To promote natural behavior and reduce task-related stress, participants were explicitly informed that there were no time constraints on their choices.
\subsection{Data analysis}
Eye-tracking data were collected and processed for both the real-world and virtual environments to ensure comparability.
In the real-world condition, participants’ gaze behavior was tracked using Tobii Pro Glasses 3. The eye-tracking data (\(t_{\text{real}} \), measured in seconds) were processed using Tobii software. Videos of participants’ eye movements were initially mapped automatically, with the system identifying gaze points and corresponding Areas of Interest (AOIs). Researchers then manually reviewed the data to correct any errors. The AOIs included both the entire supermarket shelf and individual cereal products. The time spent looking at each AOI (\( t_{\text{AOI}} \)) was measured in seconds and exported for further analysis.
In the virtual environment, eye-tracking data were collected using the Meta Quest Pro’s integrated eye-tracker, which recorded gaze behavior at a sampling rate of 90 Hz (i.e., 90 frames per second). The system tracked how long participants focused on specific AOIs by counting the number of frames during which a participant’s gaze was directed at a particular cereal product. To convert the frame data into time, the total number of frames spent viewing each cereal was divided by the 90 Hz sampling rate, yielding the time spent on each AOI in seconds.
This conversion ensured that the eye-tracking data from the virtual environment were standardized to seconds, allowing for direct comparison with the Tobii data from the real-world environment. By aligning both datasets in a common unit of measurement (\( t = s \)), we established a consistent methodology for comparing gaze behavior between the real and virtual supermarket environments.

\section{Results}
\begin{figure}[ht]
  \centering
  \includegraphics[width=\textwidth]{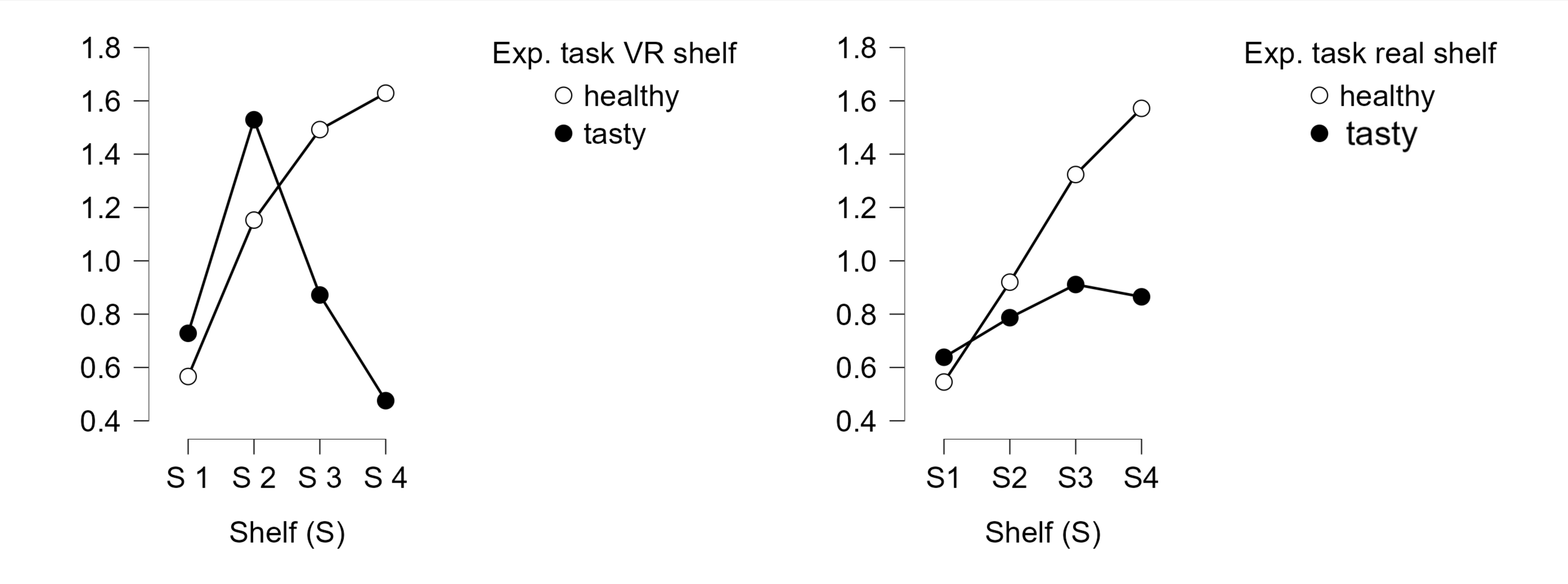}
  \caption{On the left: average viewing times per shelf in the VR environment and dependent on experimental task. On the right: average viewing times per shelf in the real environment and dependent on search task}
  \label{fixation_per_shelf}
\end{figure}

The following sections describe participants' gaze patterns on the virtual shelf and the real shelf. In particular, we were interested in the differences in gaze patterns between the two environments as indicators of differences in attention and investigated possible reasons for differences. 
First, we examined variations in overall mean fixation times across all shelves and SKUs based on experimental environments (real or VR).  Because all participants completed both experimental conditions in random order, we used a paired samples t-test to assess variation. The results indicate that the average fixation time in the real environment ($M$ = 0.90s, $SD$ = 0.65) and in the VR environment ($M$ = 1.13s, $SD$ = 0.68) are not significantly different ($t$ = -1.46; $df$ = 28, $p$ = 0.15) in our sample. The estimated effect size ($d$ = -.27, 95\% CI [-.64, .10]) indicates that the within-subject effect of the environment can be considered small-to-medium, however, with some uncertainty (\cite{kelley2012effect}).
\noindent Including experimental task in the equation indicates that differences in fixation times between environments can be explained by the experimental search tasks (sweet vs. healthy), as there is a significant interaction between search task and environment ($F$(1, 27) = 5.18, $p$ < .05) (Table \ref{tab:withinSubjectsEffects}). Specifically, mean fixation times are longer when searching for healthy products (real shelf: $M$ = 1.07s, virtual shelf $M$ = 1.19s) compared to fixation times when searching for sweet products (real shelf: $M$ = .80s, virtual shelf $M$ = .92s). 
Second, we examined search times for the four different shelf levels presented to participants (from the top shelf (S1) to the bottom shelf (S4)) (Figure \ref{fixation_per_shelf}).
\begin{table}[!ht]
	\centering
	\caption{Within-Subjects Effects for Experimental Environment (Real or VR) and Task Type (Tasty or Healthy) on Fixation Patterns.}
	\label{tab:withinSubjectsEffects}
	{
  \resizebox{\textwidth}{!}{%
		\begin{tabular}{lrrrrr}
			\hline
			Cases & Sum of Squares & df & Mean Square & F & p  \\
			\hline
			Environment (Real or VR) & $0.189$ & $1$ & $0.189$ & $0.949$ & $0.339$  \\
			Environment (Real or VR) * Experimental task (delicious or healthy) & $1.033$ & $1$ & $1.033$ & $5.175$ & $0.031$  \\
			Residuals & $5.389$ & $27$ & $0.200$ &  &  \\
			\hline
		\end{tabular}
	}}
\end{table}

\noindent Again, we examined the average fixation times for the four shelves in relation to the experimental environment (real or VR) and the experimental task (search for tasty or healthy food). 
The ANOVA results indicate that there is an interactive effect of the experimental environment and experimental task on average fixation times per shelf ($F$ (3, 81) = 7.76, $p$ < .01) (Table \ref{tab:withinSubjectsEffectsgaze}). A graphical representation of the results is shown in Figure 1. The figures indicate that when searching for healthy products, average fixation times increased when participants looked at lower shelves (Shelf (S) 4, bottom) compared to higher shelves (S1, top). For tasty products, this pattern is not visible. Instead, average fixation times remain relatively stable for all four shelves in the real environment and show a peak in average viewing times for the second shelf in the VR environment. As the reasons for this result are unclear, we will continue with an investigation of all individual SKUs.
\begin{table}[!ht]
	\centering
	\caption{Within-Subjects Effects for Shelf Position, Environment, and Task Type on Gaze Patterns.}
	\label{tab:withinSubjectsEffectsgaze}
	{
 \resizebox{\textwidth}{!}{%
		\begin{tabular}{lrrrrr}
			\hline
			Cases & Sum of Squares & df & Mean Square & F & p  \\
			\hline
			Shelf & $10.646$ & $3$ & $3.549$ & $11.608$ & $<$ .001  \\
			Shelf * Experimental task (delicious or healthy) & $1.727$ & $3$ & $0.576$ & $1.883$ & $0.139$  \\
			Residuals & $24.763$ & $81$ & $0.306$ & $$ & $$  \\
			Environment (Real or VR) & $0.665$ & $1$ & $0.665$ & $0.797$ & $0.380$  \\
			Environment (Real or VR) * Experimental task (delicious or healthy) & $4.896$ & $1$ & $4.896$ & $5.869$ & $0.022$  \\
			Residuals & $22.524$ & $27$ & $0.834$ & $$ & $$  \\
			Shelf * Environment (Real or VR) & $3.058$ & $3$ & $1.019$ & $2.162$ & $0.099$  \\
			Shelf * Environment (Real or VR) * Experimental task (delicious or healthy) & $10.983$ & $3$ & $3.661$ & $7.765$ & $<$ .001  \\
			Residuals & $38.188$ & $81$ & $0.471$ & $$ & $$  \\
			\hline
		\end{tabular}
	}}
\end{table}

\noindent Third, to investigate whether the type of SKU on each shelf might have influenced the variation in average fixation times across shelves, we examined the average fixation times per SKU for the experimental tasks (tasty or healthy). 
The descriptive statistics for fixation times per SKU in Table \ref{tab:product-shelf-position} show that participants in the VR tasty experimental task fixated longer on SKUs on shelf 2, the eye-level shelf, especially when they were not positioned as 'healthy' (the only 'healthy' product on shelf 2, Nestle Fitness Milk Chocolate, was actually fixated less in VR in the tasty experimental task compared to the healthy task). When the products were categorised as tasty or healthy, we find that around 80\% of the products were positioned as tasty and only around 20\% as healthy. We considered a product to be positioned as healthy if it used claims such as 'natural ingredients' or 'fitness' in large font on the packaging. It was not enough, for example, to display the relatively small Nutri Score. Fixation times indicate that participants were more likely to fixate on products at eye level when faced with the 'easier' task of choosing a tasty product in the less natural VR environment.

\begin{table}[!ht]
\centering
\caption{Comparison of Average Fixation Time for Products Across Experimental Tasks in Virtual Reality (VR) and Real Environments. The table displays average fixation times (M and SD, in seconds) for each product in both environments, highlighting differences in attention given to specific product types and positions. Products are organized by the shelf position (S), the positioning of the product as evaluated by the researchers (Pos) and the Nutri Score where available.}
\label{tab:product-shelf-position}
\resizebox{\textwidth}{!}{%
\begin{tabular}{lccccccccccc}
\textit{} & \multicolumn{1}{l}{\textit{}} & \multicolumn{1}{l}{\textit{}} & \multicolumn{1}{l}{\textit{}} & \multicolumn{4}{c}{\textit{\textbf{\begin{tabular}[c]{@{}c@{}}Experimental task: \\ Healthy\end{tabular}}}} & \multicolumn{4}{c}{\textit{\textbf{\begin{tabular}[c]{@{}c@{}}Experimental task: \\ Tasty\end{tabular}}}} \\ \hline
\textit{\textbf{Product Name}} & \textit{\textbf{\begin{tabular}[c]{@{}c@{}}S\end{tabular}}} & \textit{\textbf{Pos }} & \textit{\textbf{\begin{tabular}[c]{@{}c@{}}Nutri  \\ Score\end{tabular}}} & \multicolumn{2}{c}{\textit{\textbf{\begin{tabular}[c]{@{}c@{}}VR \\ (N = 21)\end{tabular}}}} & \multicolumn{2}{c}{\textit{\textbf{\begin{tabular}[c]{@{}c@{}}Real \\ (N = 12)\end{tabular}}}} & \multicolumn{2}{c}{\textit{\textbf{\begin{tabular}[c]{@{}c@{}}VR \\ (N = 14)\end{tabular}}}} & \multicolumn{2}{c}{\textit{\textbf{\begin{tabular}[c]{@{}c@{}}Real \\ (N = 20)\end{tabular}}}} \\ \hline
\multicolumn{1}{c}{} &  &  &  & M & SD & M & SD & M & SD & M & SD \\ \hline
Happy   Hoops (Gut\&Günstig) & 1 & tasty &  & 0.68 & 0.81 & 0.55 & 0.75 & 1.21 & 2.46 & 0.59 & 0.68 \\
Honey   Wheat (Kornmühle) & 1 & tasty & C & 0.87 & 1.02 & 0.82 & 1.42 & 1.56 & 3.23 & 0.83 & 1.04 \\
Leo Crisp   (Gut\&Günstig) & 1 & tasty &  & 0.72 & 1.26 & 0.53 & 0.73 & 0.39 & 0.39 & 0.42 & 0.57 \\
Nesquik   Intense choco (Nestle) & 1 & tasty & A & 0.44 & 0.58 & 0.27 & 0.32 & 0.91 & 1.12 & 0.78 & 0.94 \\
Nesquik   Maxi Pack (Nestle) & 1 & tasty & A & 0.36 & 0.29 & 0.66 & 0.72 & 0.82 & 0.83 & 0.88 & 1.03 \\
Nougat   Bits (Gut\&Günstig) & 1 & tasty &  & 0.57 & 0.45 & 0.53 & 0.46 & 0.76 & 0.95 & 0.66 & 0.99 \\
Cookie   Crips (Nestle) & 2 & tasty & C & 0.68 & 0.46 & 0.57 & 0.32 & \textbf{1.62} & 1.32 & 0.45 & 0.66 \\
Fitness   Milk chocolate (Nestle) & 2 & healthy & C & 2.85 & 3.04 & 1.87 & 3.46 & \textbf{1.98} & 2.06 & 0.80 & 0.88 \\
Honey   Wheat (Gut\&Günstig) & 2 & tasty &  & 1.47 & 1.58 & 0.96 & 0.71 & 2.81 & 3.93 & 1.01 & 1.24 \\
Crunchy   Nut Bites (Kellogg's) & 2 & tasty &  & 0.69 & 0.75 & 0.76 & 0.66 & 0.67 & 0.41 & 0.98 & 0.84 \\
Tresor   choco, caramel (Kellogg's) & 2 & tasty &  & 0.67 & 1.02 & 0.48 & 0.41 & 1.61 & 4.44 & 0.53 & 0.57 \\
Tresor   dark chocolate (Kellogg's) & 2 & tasty &  & 0.68 & 0.76 & 0.76 & 0.94 & 1.12 & 1.30 & 1.04 & 1.48 \\
Lion   triple crunchy (Nestle) & 2 & tasty & D & 0.51 & 0.47 & 0.67 & 0.60 & 1.02 & 1.07 & 0.85 & 0.66 \\
Clusters   (Nestle) & 3 & tasty & C & 0.91 & 0.92 & 1.36 & 2.24 & 0.55 & 0.55 & 0.55 & 0.72 \\
KitKat   cereal (KitKat) & 3 & tasty & C & 0.57 & 0.58 & 0.32 & 0.30 & 1.36 & 1.15 & 0.81 & 0.94 \\
Lion   Caramel Chocolat (Nestle) & 3 & tasty & C & 0.79 & 0.80 & 0.84 & 0.54 & 1.37 & 1.18 & 0.87 & 0.94 \\
Oreo O's   cereal (Oreo) & 3 & tasty &  & 0.49 & 0.37 & 0.28 & 0.50 & 2.14 & 2.67 & 0.54 & 0.70 \\
Chillo   Pillows (Wholey) & 3 & healthy &  & 1.56 & 2.06 & 1.69 & 1.67 & 0.91 & 1.15 & 1.12 & 1.90 \\
Cinna   Rollies (Wholey) & 3 & healthy &  & 1.52 & 1.52 & 2.01 & 3.15 & 1.05 & 2.21 & 1.10 & 1.82 \\
Lucky   Loops (Wholey) & 3 & healthy &  & 3.39 & 3.14 & 2.49 & 2.29 & 1.37 & 2.73 & 1.45 & 1.25 \\
Cheerios   (Nestle) & 4 & tasty & B & 1.22 & 1.25 & 2.76 & 4.26 & 0.59 & 0.87 & 0.76 & 1.11 \\
Cini Minis   Churros (Nestle) & 4 & tasty & C & 1.15 & 0.94 & 1.06 & 1.01 & 0.57 & 0.71 & 1.07 & 0.86 \\
Flakes   (Kaufland classic) & 4 & tasty &  & 2.72 & 3.09 & 1.88 & 2.72 & 0.95 & 1.69 & 0.89 & 1.05 \\
Fitness  Dark Chocolat (Nestle) & 4 & healthy & B & 2.74 & 4.05 & 1.04 & 1.23 & 0.66 & 1.56 & 0.99 & 1.36 \\
Nesquik   Bio (Nestle) & 4 & tasty & A & 0.98 & 1.16 & 0.75 & 0.79 & 0.74 & 0.90 & 0.58 & 0.70 \\
Nougat   Bits (Granola) & 4 & tasty & D & 1.10 & 1.23 & 1.35 & 1.67 & 0.83 & 0.95 & 0.93 & 1.26 \\ \hline
 & \multicolumn{1}{l}{} & \multicolumn{1}{l}{} & \multicolumn{1}{l}{} & \multicolumn{1}{l}{} & \multicolumn{1}{l}{} & \multicolumn{1}{l}{} & \multicolumn{1}{l}{} & \multicolumn{1}{l}{} & \multicolumn{1}{l}{} & \multicolumn{1}{l}{} & \multicolumn{1}{l}{} \\
 & \multicolumn{1}{l}{} & \multicolumn{1}{l}{} & \multicolumn{1}{l}{} & \multicolumn{1}{l}{} & \multicolumn{1}{l}{} & \multicolumn{1}{l}{} & \multicolumn{1}{l}{} & \multicolumn{1}{l}{} & \multicolumn{1}{l}{} & \multicolumn{1}{l}{} & \multicolumn{1}{l}{} \\
 & \multicolumn{1}{l}{} & \multicolumn{1}{l}{} & \multicolumn{1}{l}{} & \multicolumn{1}{l}{} & \multicolumn{1}{l}{} & \multicolumn{1}{l}{} & \multicolumn{1}{l}{} & \multicolumn{1}{l}{} & \multicolumn{1}{l}{} & \multicolumn{1}{l}{} & \multicolumn{1}{l}{} \\
 & \multicolumn{1}{l}{} & \multicolumn{1}{l}{} & \multicolumn{1}{l}{} & \multicolumn{1}{l}{} & \multicolumn{1}{l}{} & \multicolumn{1}{l}{} & \multicolumn{1}{l}{} & \multicolumn{1}{l}{} & \multicolumn{1}{l}{} & \multicolumn{1}{l}{} & \multicolumn{1}{l}{} \\
 & \multicolumn{1}{l}{} & \multicolumn{1}{l}{} & \multicolumn{1}{l}{} & \multicolumn{1}{l}{} & \multicolumn{1}{l}{} & \multicolumn{1}{l}{} & \multicolumn{1}{l}{} & \multicolumn{1}{l}{} & \multicolumn{1}{l}{} & \multicolumn{1}{l}{} & \multicolumn{1}{l}{} \\
 & \multicolumn{1}{l}{} & \multicolumn{1}{l}{} & \multicolumn{1}{l}{} & \multicolumn{1}{l}{} & \multicolumn{1}{l}{} & \multicolumn{1}{l}{} & \multicolumn{1}{l}{} & \multicolumn{1}{l}{} & \multicolumn{1}{l}{} & \multicolumn{1}{l}{} & \multicolumn{1}{l}{} \\
 & \multicolumn{1}{l}{} & \multicolumn{1}{l}{} & \multicolumn{1}{l}{} & \multicolumn{1}{l}{} & \multicolumn{1}{l}{} & \multicolumn{1}{l}{} & \multicolumn{1}{l}{} & \multicolumn{1}{l}{} & \multicolumn{1}{l}{} & \multicolumn{1}{l}{} & \multicolumn{1}{l}{} \\
 & \multicolumn{1}{l}{} & \multicolumn{1}{l}{} & \multicolumn{1}{l}{} & \multicolumn{1}{l}{} & \multicolumn{1}{l}{} & \multicolumn{1}{l}{} & \multicolumn{1}{l}{} & \multicolumn{1}{l}{} & \multicolumn{1}{l}{} & \multicolumn{1}{l}{} & \multicolumn{1}{l}{}
\end{tabular}%
}
\end{table}
\section{Discussion}
Our results indicate that, overall, attention patterns in the virtual environment and the real supermarket shelf were similar, as no significant differences in gaze behavior were found between VR and the real shelf. This finding aligns with previous research by Branca et al. \cite{Branca2023-yf} and Siegrist et al. \cite{Siegrist2019-yo}. However, a closer examination of the shelves reveals notable differences between the two experimental scenarios.
One key difference emerged when comparing gaze patterns across the four shelf levels in the two experimental tasks (tasty and healthy product selections). For healthy products, participants spent more time looking at products located on the lower shelves, whereas for tasty products, they spent relatively little time focusing on the lower shelves.
In the real environment, participants spent less time looking at all tasty products overall, whereas in the VR environment, they spent more time focusing on products located on the second shelf (eye level). This pattern in the real-world setting seems reasonable, as identifying tasty products on a cereal shelf may be easier than identifying healthy ones. Many tasty cereals prominently display chocolate, honey, or other sweet ingredients, making them visually distinct and easy to recognize. In contrast, healthy cereals often feature subtle health-related information that may be difficult to notice. Health claims, such as “X\% less added sugar” or the Nutri-Score, are often displayed in small print on the front of the packaging, sometimes on the lowest shelf (shelf 4), making them harder to read. Furthermore, Nutri-Scores may not always align with consumer expectations based on product placement; for example, some cereals with a Nutri-Score of A may not explicitly highlight their health benefits on the front of the pack.
Despite these findings, participants in the VR environment spent significant time on the second shelf, which is at eye level—a shelf height generally considered optimal for consumer attention and product selection. This suggests that while overall attention patterns may be similar in real and virtual environments, specific gaze behaviors—particularly when focusing on individual shelves or products—can differ between the two settings.
Several factors could explain these discrepancies between VR and real-world gaze patterns, as this result contrasts with prior studies (\cite{Siegrist2019-yo}).
One potential reason is the nature of the experimental task and the VR interaction constraints. The VR shopping environment lacked full product interactivity, meaning participants could not interact with products in the usual way—such as picking them up or examining detailed packaging information. As a result, participants may have opted for a simpler, more surface-level search strategy, focusing only on products that were easiest to find visually.
Another explanation may stem from how we analyzed our data, as we focused on gaze patterns rather than actual product choices. Prior research suggests that decision-making in shopping contexts often follows simple heuristics or decision rules (\cite{kalnikaite2013decision}). While product choices may be relatively consistent across environments, gaze behavior could be influenced by the specific laboratory setup and differences in how information was presented in each condition.
A third factor concerns the amount of contextual information available to participants. In both conditions, the experimental shelves lacked typical supermarket cues, such as advertising materials, brand promotions, and shelf organization based on consumer expectations. Most critically, prices were omitted to simplify the scenario. However, pricing is a major factor in consumer decision-making (\cite{dickson1990price}), and its absence may have altered how participants searched for and evaluated products in both environments.

\section{Conclusion}
Our findings suggest that while VR can replicate general shopping patterns, it may be premature to conclude that consumer behavior in VR fully mirrors real-world shopping behavior (\cite{Branca2023-yf,Branca2024-ef}). However, as discussed above, certain limitations in our study design may have influenced the results.
Future research should further explore whether VR shopping environments are perceived as more challenging than real-world settings, as well as which specific factors contribute to this perception. One avenue of research could investigate different VR technologies and presentation styles, as our study used a VR environment closely modeled after the real supermarket setup. Alternative VR presentations could potentially enhance the shopping experience, for example, by providing additional digital product information or offering alternative product organization methods (\cite{Branca2024-ef}). The effectiveness of such adaptations may depend on the type of product being studied—for example, fast-moving consumer goods like groceries may require different VR adaptations than durable goods such as electronics or furniture.
Another important area for future research would be the inclusion of additional contextual factors, particularly price information, to make VR shopping experiences more comparable to real-world settings. Finally, future studies could expand the range of dependent variables beyond gaze patterns to explore decision-making strategies, purchase behavior, and user experience in greater depth.
\begin{credits}
\subsubsection{\ackname} During the preparation of this work, the author(s) used X-GPT-4 in order to: Grammar and spelling check. After using these tool(s)/service(s), the author(s) reviewed and edited the content as needed and take(s) full responsibility for the publication’s content. 
\end{credits}
%
%
%
 \bibliographystyle{splncs04}
 \bibliography{main}

\end{document}